\documentclass[twoside,twocolumn,9pt]{article}
\usepackage{extsizes}
\usepackage[super,sort&compress,comma]{natbib} 
\usepackage[version=3]{mhchem}
\usepackage[left=1.5cm, right=1.5cm, top=1.785cm, bottom=2.0cm]{geometry}
\usepackage{balance}
\usepackage{widetext}
\usepackage{times,mathptmx}
\usepackage{sectsty}
\usepackage{graphicx} 
\usepackage{lastpage}
\usepackage[format=plain,justification=raggedright,singlelinecheck=false,font={stretch=1.125,small,sf},labelfont=bf,labelsep=space]{caption}
\usepackage{float}
\usepackage{fancyhdr}
\usepackage{fnpos}
\usepackage[english]{babel}
\usepackage{array}
\usepackage{droidsans}
\usepackage{charter}
\usepackage[T1]{fontenc}
\usepackage[usenames,dvipsnames]{xcolor}
\usepackage{setspace}
\usepackage[compact]{titlesec}
\usepackage{wasysym}
\usepackage{epstopdf}

\begin{document}

\pagestyle{fancy}
\thispagestyle{plain}

\makeFNbottom
\makeatletter
\renewcommand\LARGE{\@setfontsize\LARGE{15pt}{17}}
\renewcommand\Large{\@setfontsize\Large{12pt}{14}}
\renewcommand\large{\@setfontsize\large{10pt}{12}}
\renewcommand\footnotesize{\@setfontsize\footnotesize{7pt}{10}}
\makeatother

\renewcommand{\thefootnote}{\fnsymbol{footnote}}
\renewcommand\footnoterule{\vspace*{1pt}%
\color{black}\hrule width 3.5in height 0.4pt \color{black}\vspace*{5pt}} 
\setcounter{secnumdepth}{5}

\makeatletter 
\renewcommand\@biblabel[1]{#1}            
\renewcommand\@makefntext[1]%
{\noindent\makebox[0pt][r]{\@thefnmark\,}#1}
\makeatother 
\renewcommand{\figurename}{\small{Fig.}~}
\sectionfont{\sffamily\Large}
\subsectionfont{\normalsize}
\subsubsectionfont{\bf}
\setstretch{1.125} 
\setlength{\skip\footins}{0.8cm}
\setlength{\footnotesep}{0.25cm}
\setlength{\jot}{10pt}
\titlespacing*{\section}{0pt}{4pt}{4pt}
\titlespacing*{\subsection}{0pt}{15pt}{1pt}

\fancyhead{}
\renewcommand{\headrulewidth}{0pt} 
\renewcommand{\footrulewidth}{0pt}
\setlength{\arrayrulewidth}{1pt}
\setlength{\columnsep}{6.5mm}
\setlength\bibsep{1pt}

\makeatletter 
\newlength{\figrulesep} 
\setlength{\figrulesep}{0.5\textfloatsep} 

\makeatother

\twocolumn[
  \begin{@twocolumnfalse}
\vspace{3cm}
\sffamily

\noindent\LARGE{\textbf{Temperature dependence of the hydrogen bond network in Trimethylamine N-oxide and guanidine hydrochloride -- water solutions}} \\
\vspace{0.3cm} \\ 

  \noindent\large{Felix Lehmk\"uhler,$^{\ast}$\textit{$^{a,b}$} Yury Forov,\textit{$^{c}$} Mirko Elbers,\textit{$^{c}$} Ingo Steinke,\textit{$^{a,b}$} Christoph J. Sahle,\textit{$^{d}$} Christopher Weis,\textit{$^{c}$} Naruki Tsuji,\textit{$^{e}$} Masayoshi Itou,\textit{$^{e}$} Yoshiharu Sakurai,\textit{$^{e}$} Agnieszka Poulain,\textit{$^{d}$} and Christian Sternemann\textit{$^{c}$}} \\

 \noindent\normalsize{We present an X-ray Compton scattering study on aqueous Trimethylamine N-oxide (TMAO) and guanidine hydrochloride solutions (GdnHCl) as a function of temperature. Independent from the concentration of the solvent, Compton profiles almost resemble results for liquid water as a function of temperature. However, The number of hydrogen bonds per water molecule extracted from the Compton profiles suggests a decrease of hydrogen bonds with rising temperatures for all studied samples, the differences between water and the solutions are weak. Nevertheless, the data indicate a reduced bond weakening with rising TMAO concentration up to 5M of 7.2\%\ compared to 8\%\ for pure water. In contrast, the addition of GdnHCl appears to behave differently for concentrations up to 3.1 M with a weaker impact on the temperature response of the hydrogen bond structure.} \\

 \end{@twocolumnfalse} \vspace{0.6cm}

  ]

\renewcommand*\rmdefault{bch}\normalfont\upshape
\rmfamily
\section*{}
\vspace{-1cm}

\footnotetext{\textit{$^{a}$~Deutsches Elektronen-Synchrotron DESY, Notkestr. 85, 22607 Hamburg, Germany; Fax: +49 40 8998 2787; Tel: +49 40 8998 5671; E-mail: felix.lehmkuehler@desy.de}}
\footnotetext{\textit{$^{b}$~The Hamburg Centre for Ultrafast Imaging, Luruper Chaussee 149, 22761 Hamburg, Germany.}}
\footnotetext{\textit{$^{c}$~Fakult\"at Physik/DELTA, Technische Universit\"at Dortmund, 44221 Dortmund, Germany.}}
\footnotetext{\textit{$^{d}$~ESRF -- The European Synchrotron, CS 40220, 38043 Grenoble Cedex 9, France.}}
\footnotetext{\textit{$^{e}$~SPring-8, JASRI, 1-1-1 Kuoto, Sayo-gun, Hyogo 679-5198, Japan}}

\section{Introduction}
The complexity of the hydrogen bond network of liquid water is presumed to be responsible for water's properties, especially the occurrence of the various water anomalies \cite{debenedetti2003,nilsson2011,nilsson2015,Perakis2017}. Although its structure has frequently been addressed in many experiments for decades, many questions are still unsolved. In particular, an impact from so-called chaotrop and cosmotropic solutes on the structure of liquid water is widely accepted, but direct experimental proof is still scarce. While chaotrop substances disturb the hydrogen bond network upon solution, cosmotropic materials act as structure former, i.e., they are believed to strengthen the network. A frequently studied example for a chaotropic material is guanidine hydrocloride (CH$_6$ClN$_3$, GdnHCl), while trimethylamine N-oxide (C$_3$H$_9$NO, TMAO) represents a cosmotropic specimen. Both are known to denaturate (GdnHCl) \cite{Bhuyan2002} or stabilize (TMAO) \cite{Ma2014} proteins in water, respectively. These processes are believed to be mediated indirectly via the impact of the solutes on the water structure. However, recent work questioned these assumptions \cite{Batchelor2004}, suggesting a weakening of water hydrogen bonds in presence of both species \cite{Pazos2012,Knake2015}. Especially the influence of TMAO on the water network is the objetct of various experimental and simulation work. Therein, contradictory results are found. Mid-infrared pump-probe spectroscopy reported on increasing densities of network defects in the hydrogen-bond network of water \cite{Rezus2009}. In contrast, dielectric and Raman spectroscopy combined with electronic structure computations suggests that the oxygen atom in TMAO accepts on average at least three hydrogen bonds from neighboring water molecules \cite{Munroe2011}. A further study reported that stable TMAO-H$_2$O complexes are assumed to form incorporating two to three water molecules per TMAO molecule \cite{Hunger2012}. Furthermore, the influence of TMAO was studied by X-ray Raman scattering, suggesting a more structured hydrogen-bond network in the presence of TMAO \cite{Sahle2016}. Consequently, the question of how and if on average the addition of TMAO modifies the number of hydrogen bonds per water molecule in water-TMAO solutions remains unsolved.

In liquid water, the average number of hydrogen bonds per molecule decreases with increasing temperature \cite{hakala2006,Rastogi2011,sahle2013}. Although the impact of TMAO and GdnHCl on the structure of water is in the focus of various studies, little is known about how the cosmotropic TMAO and the chaotropic GdnHCl stabilize or disturb the hydrogen bond geometry at different temperatures. In particular, a potential counteraction of hydrogen bond strengthening or weakening with temperature and addition of TMAO as well as GdnHCl are unknown so far. This can be accessed by means of X-ray Compton scattering as demonstrated by earlier studies on water \cite{hakala2006,sit2007} and aqueous solutions \cite{lehmkuehler2011}.

In this study, we use Compton scattering to evaluate the impact of TMAO and GdnHCl on the hydrogen bond network of water as a function of temperature. Aqueous suspensions are studied at molar ratios between 0 M and 5 M over a temperature range between 273 K and 333 K. For all studied systems, a decreasing average number of hydrogen bonds per molecule is observed with rising temperature. However, no significant variations of Compton profile differences between the solutions and liquid water can be found, suggesting that aqueous solutions of cosmotropes and chaotropes show a similar structural response upon increasing temperature.

\section{Experimental}
\subsection{Compton scattering}
In recent years, non-resonant inelastic X-ray scattering has become a standard technique to investigate liquid samples \cite{isaacs1999,bellin2011,juurinen2011,juurinen2013,juurinen2014b,sahle2013,lehmkuehler2016,Sahle2016}. In the limit of large energy and momentum transfer, Compton scattering dominates and the impulse approximation  becomes valid \cite{Cooper2004,schuelke}. For liquid samples, the measured quantity in such Compton scattering experiments is proportional to the so-called Compton profile for isotropic systems
\begin{equation}
J(p_q)=\frac{1}{2}\int\text{d}\Omega\int_{|p_q|}^\infty \rho(\textbf{p})p\text{d}p.
\end{equation}
Here $p_q$ denotes a scalar electron momentum variable. Compton profiles are normalized to the number of electrons per molecule \cite{Cooper2004}. Obviously, the Compton profile is related to the ground state electron momentum density $\rho(\textbf{p})$. Changes of ground state can be quantified by the integral over the normalized Compton profile
\begin{equation}
n=\int\limits_0^\infty |\Delta J(p_q)|\mathrm{d}p_q.
\label{eq:integ}
\end{equation} 
The value of $n$ is interpreted as a measure of the number of electrons whose wave function changes and provides in particular quantitative information for changes of the sample's bond geometry. Comparing computed Compton differences based on structural snapshots with experimental data the inverse value of $n$ was found to be linearly related to the number of hydrogen bonds per water molecule in liquid and supercritical water \cite{sit2007}. This quantity bas been used so far to discuss Compton profiles of ionic liquids \cite{koskelo} and hydrogen bonds in ice \cite{bellin2011}.

In recent publications, the Compton profile was demonstrated to be very sensitive to single particle properties and small changes in the intra- and intermolecular bond geometry in molecular systems \cite{juurinen2014}. A large number of studies concentrated on water-based systems such as liquid, confined, supercooled and supercritical water \cite{hakala2006,hakala2006b,nygard2007,sit2007,reiter2013,lehmkuehler2016,reiter2016}, structure and energetics of ice \cite{isaacs1999,ragot2002,nygardprl,bellin2011} and two-component systems \cite{sternemann2006,hakala2009,lehmkuehler2010,lehmkuehler2011,juurinen2011} and demonstrated the power of Compton scattering to uniquely probe local structures and energetics in hydrogen-bonded systems. In particular, by matching experimental data with theory the particular sensitivity of Compton scattering to quantum effects on the hydrogen and covalent bonds was demonstrated \cite{nygard2007,sit2007,lehmkuehler2010,lehmkuehler2016}.

\subsection{Experimental Setups}
The experiments have been performed at beamline ID15B of the European Synchrotron Radiation Facility (ESRF) \cite{Hiraoka2005} and beamline BL08W at SPring-8 \cite{Hiraoka2001}. The energy of the incident X-ray beam was 87.17 keV (ESRF) and 182.7 keV (SPring-8). The beam size was set to $0.5\times0.5$ mm defined by slits. The scattered intensity was measured by multi-element Ge solid-state detectors in backscattering geometry. We used 13 elements at a scattering angle of 150$^\circ$ at ESRF and 10 elements at an angle of 170$^\circ$ at SPring-8. Momentum resolutions of $\Delta p_q\approx 0.65$ atomic units (a.u.) at ESRF and $\Delta p_q\approx 0.57$ a.u.~at Spring-8 were achieved at the Compton peak ($p_q=0$ a.u.). Similar to previous experiments, the incident flux at the ESRF was kept constant using an absorber feedback system to achieve constant detector conditions. At Spring-8, the incoming intensity is constant due to the top-up operation of the storage ring. The obtained statistical accuracy was better than 0.025 \% units at $p_q=0$ a.u.~within 0.03 a.u.~momentum bin in both setups.

The samples were filled into glass capillaries of 2 mm thickness that were sealed afterwards. First, ultra-pure water (milli-Q, $R>18$ M$\Omega$) was measured as reference. In order to investigate the solutes' influence on the water structure and neglect solute-solute interactions, we decided to use moderately concentrated solutions. For TMAO, we chose concentrations of 0.8 M, 3 M, and 5 M, for guanidine hydrochloride two concentrations were prepared with 1.1 M and 3.1 M, respectively. The capillaries were placed into a sample holder that was capable to cover a temperature range between 270 K and 330 K with high statistical accuracy of better than $\Delta T=20$ mK within several hours of experimental time. The samples were measured at different temperatures between 273 K and 333 K, typically in steps of 10 to 20 K.

For consistency, Compton spectra were saved every 10 minutes and checked for deviations larger than the statistical accuracy. In total, spectra were measured for four to eight hours per temperature. The raw spectra data were corrected for absorption and the dead times of the detector before converting to momentum scale by using the relativistic cross section correction \cite{Cooper2004}. Contributions from multiple scattering were corrected afterwards using a recent Monte Carlo code \cite{montecarlo}. In order to neglect any background contribution, typically profile differences are studied rather than pure Compton profiles. However, two different sample systems typically cannot be compared directly because of density differences or small set-up deviations, e.g.~different sample thicknesses, in our case small deviations for two capillaries, lead to significant differences of the Compton profile that cannot be corrected for. Hence, we have to limit ourselves to temperature-induced changes within the sample systems. We calculated the profile differences with respect to the data at 273 K for each detector element separately before averaging over all elements. Finally, the positive and negative momentum sides of the Compton profile differences were averaged.

\section{Results and Discussion}
\begin{figure}
	\includegraphics[width=\columnwidth]{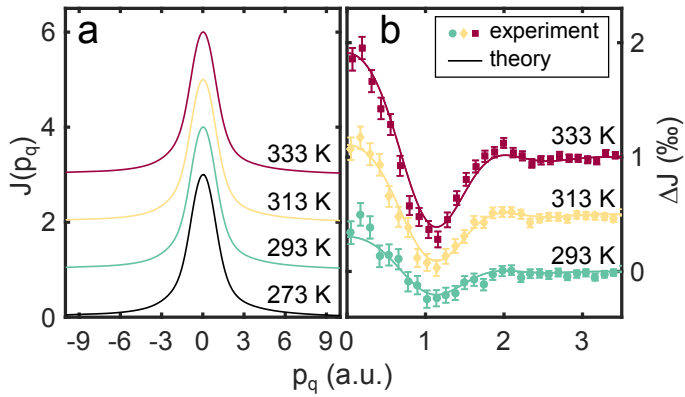}%
	\caption{(a) Compton profiles of water at different temperatures as indicated. (b) Compton profile differences with respect to 273 K. The line is a scaled difference based on DFT calculations. Data are shifted vertically for clarity.}%
	\label{fig1}%
\end{figure}

First, we focus on changes of the Compton profile of liquid water as a function of temperature. Compton profiles measured at four temperatures between 273 K and 333 K are shown in Fig.~\ref{fig1} (a). 

As discussed above, variations of the profiles are weak. This is highlighted by calculating the profile differences with respect to 273 K, i.e.
\begin{equation}
\Delta J(p_q)=\frac{J_T(p_q)-J_\text{273}(p_q)}{J_\text{273}(0)}
\end{equation}
shown in Fig.~\ref{fig1} (b). All differences show a pronounced maximum around $p_q=0$ a.u.~accompanied by a minimum at $p_q\approx1.1$ a.u.~whose amplitudes increase with increasing temperature. We find a maximum difference of about $\Delta J(p_q=0)=1$~\permil\ for $\Delta T=60$ K. Such a shape has been reported frequently as a typical result for temperature-induced changes of the hydrogen bond network in liquid water and hexagonal ice \cite{hakala2006,hakala2006b,nygardprl}. In this way it provides a fingerprint of the weakening of hydrogen bonds with rising temperature. The experimental data is compared to calculations in the framework of density functional theory (DFT), using structural snapshots of pure liquid water extracted from ab-initio molecular dynamics (MD) simulations \cite{nygard2007}. The DFT data shown as lines in Fig.~\ref{fig1} are scaled linearly with respect to the corresponding temperature difference. As expected, we do not observe any significant variations between experiment, DFT data and thus to earlier experimental differences \cite{hakala2006,nygard2007}, proving stable experimental conditions.

\begin{figure}
	\includegraphics[width=\columnwidth]{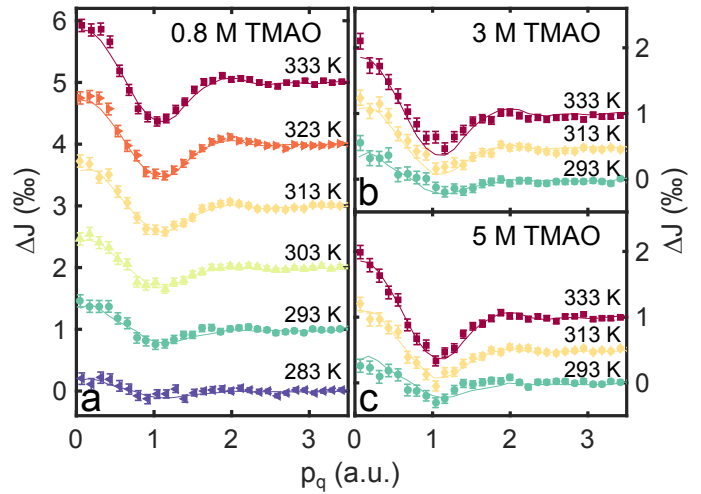}%
	\caption{Compton profile differences of water-TMAO solutions at different TMAO concentrations: (a) 0.8 M TMAO, (b) 3 M TMAO, and (c) 5 M TMAO. The solid lines are smoothed and scaled water differences from Fig.~\ref{fig1} (b). Data are shifted vertically for clarity, the $\Delta J$ scale is the same for all subpanels.}%
	\label{fig2}%
\end{figure}

To determine the effect of TMAO on the hydrogen bond network of water, we measured Compton profiles of water-TMAO solutions as a function of temperature between 273 K and 333 K. Compton profile differences with respect to the Compton profile measured at 273 K are shown in Fig.~\ref{fig2}. In general, the amplitude and shape of the differences resemble fully the results for pure water. This is highlighted by the solid lines that correspond to smoothed differences of water from Fig.~\ref{fig1} (b). If the corresponding temperature difference has not been measured for water, the difference has been scaled linearly with $\Delta T$. Remarkably, the effect of TMAO concentration is weak.

\begin{figure}
	\includegraphics[width=\columnwidth]{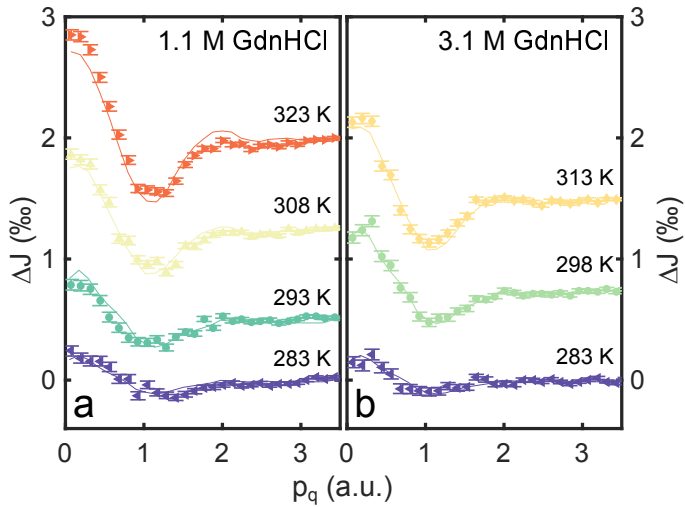}%
	\caption{Compton profile differences of water-GdnHCl solutions at different GdnHCl concentrations: (a) 1.1 M GdnHCl and (b) 3.1 M GdnHCl. The solid lines are smoothed and scaled water differences from Fig.~\ref{fig1} (b). Data are shifted vertically for clarity.}%
	\label{fig3}%
\end{figure}

The result for GdnHCl solutions are shown in Fig.~\ref{fig3}. Compared to the TMAO results, a smaller temperature range could be covered due to experimental limitations. Small deviations can be observed for the largest temperature difference only for 1.1 M GdnHCl, i.e.~a slightly stronger peak around $p_q=0$ a.u.~and accordingly less contribution at the second local maximum at $p_q\approx2$ a.u. However, this cannot be observed for other temperatures and concentrations, the water difference mainly lies within the error bars of the experimental data. Most importantly, shape and amplitude match the results for liquid water, especially the position of the minimum around $p_q=1.1$ a.u.

The differences suggest that both the addition of TMAO or GdnHCl as prototypical kosmotrope and chaotrop do not significantly change the response of temperature of the hydrogen bond network compared to pure water. Taking the conventional understanding of hydrogen bonded water into account, an increasing (decreasing) strength of the hydrogen bond, e.g.~reflected in a shortening of the hydrogen bond length, is accompanied by a stretched (compressed) average intramolecular covalent OH bond length. Thus, the addition of solvents may result in modifications of the Compton profile beyond the observed water-like differences, such as OH bond length or bond angle variations \cite{hakala2006b}. These mostly intramolecular changes manifest typically in contributions at larger $p_q$ as reported as the dominating effect of deuteration \cite{nygard2007}, in water-ethanol mixtures \cite{juurinen2011}, during freezing of clathrate hydrates \cite{lehmkuehler2010}, and in confined and supercooled water \cite{reiter2013,lehmkuehler2016}. However, such modifications are absent at first view.

\begin{figure}
	\includegraphics[width=\columnwidth]{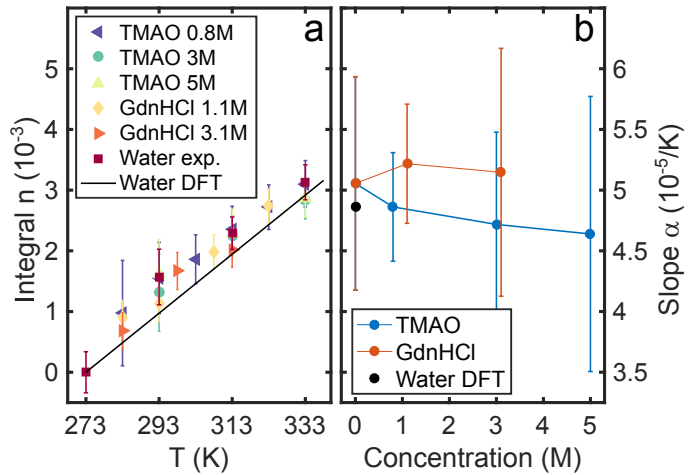}
	\caption{(a) Integrals $n$ for all measured sample systems compared to DFT results for liquid water \cite{nygard2007}. (b) Slope $\alpha$ representing the temperature dependence for each sample.}
	\label{fig4}
\end{figure}

For a more direct and qualitative comparison of the samples, we calculate the integral $n$ from Eq.~\ref{eq:integ} for each difference. Since we do not observe any indication for significant intramolecular changes, $n$ can provide a measure of the number of hydrogen bonds per water molecule. Here, we focus on changes with respect to $T=273$ K. The results are shown in Fig.~\ref{fig4} (a). The line represents the linear change for the water profiles calculated within DFT. Remarkably, the experimental data of TMAO and GdnHCl also increases linearly with $T$ and thus show only small deviations from the DFT results for liquid water. Furthermore, they match the experimental integrals of liquid water well. 

Sit et al.~\cite{sit2007} found a linear relation between $n$ and the number of hydrogen bonds $n_\text{HB}$ per molecule for water. Applying this relation to our results, we find first for the maximum $n$ of $n_\text{max}=(3.1\pm0.4)\times 10^{-3}$ a reduction of hydrogen bonds per molecule of about $8$\%\ for the temperature difference of $\Delta T=60$ K in water. Second, the statistical accuracy of the experiments results to an error for $n$ of about 1\%, i.e., for liquid water at ambient conditions changes can be determined with a high accuracy of $\delta n_\text{HB}\approx 0.04$. However, since we have only access to relative changes in Compton scattering experiments, we determined the slope for all studied samples by modelling a linear change of $n$ with slope $\alpha$. The results are shown in Fig. \ref{fig4} (b). Therein, the experimental results for liquid water are used as data at 0 M solvent concentration. The DFT data results to a similar value marked by the black point. In general, the slopes of $n$ of the TMAO and GdnHCl solutions resemble the results for water within the error bars. The obtained values around $\alpha\approx 5\cdot10^{-5}$/K suggests a reduction of $n_\text{HB}$ of less than 0.15\%\ per 1 K heating for all samples. In addition, we can observe trends in the behaviour of TMAO and GdnHCl solutions. A higher concentration of TMAO reduces the slope of $n(T)$ slightly. Thus, the addition of TMAO is suggested to counteract the thermally induced disorder of the hydrogen bond network. However, the effects are very weak, the slope drops from $\alpha(\text{0M})=5.1\cdot 10^{-5}$/K for water to $\alpha(\text{5M})=4.6\cdot 10^{-5}$/K, with overlapping error bars. This corresponds to a reduction of hydrogen bonds per molecule of (7.2 $\pm$ 1.6)\%\ for 5M TMAO and therefore slightly below the values obtained for pure water of (8 $\pm$ 1)\%. In contrast, it can be inferred that increasing the GdnHCl concentration hardly affects the slope.

\section{Conclusions}
We presented a Compton scattering study on the hydrogen bond geometry in aqueous TMAO and GdnHCl solutions hydrogen bond geometry as a function of temperature. Compton profile differences of both sample systems with respect to 273 K show the typical shape reported in various studies on aqueous systems which are understood as a fingerprint hydrogen bond weakening at elevated temperatures. Apart from the shape, the amplitude of the Compton profile differences matches experimental and DFT results for liquid water, independent from the concentration of the solvent. Analysing the number of hydrogen bonds per water molecule, we found similar results for all samples with slight indications of bond strengthening due to the presence of TMAO up to 5 M in water, counteracting the effect of increased temperature on the microscopic structure. In contrast, the addition of GdnHCl up to 3.1 M appears to have no significant impact on the temperature response of the hydrogen bond structure. However, the observed differences are very weak and close to the detection limit of state-of-the art experiments. Our results indicate that there are weak influences of TMAO and GdnHCl on the hydrogen bond network of water at different temperatures -- if there are any. This requests simulation studies on the microscopic structure of aqueous systems, in particular, taking quantum effects into account as typically measured in Compton scattering experiments.

\section*{Acknowledgements}
This work has been supported by the Cluster of Excellence "The Hamburg Centre for Ultrafast Imaging" (CUI) funded by DFG (EXC 1074). The Compton scattering experiments at SPring-8 were performed with the approval of JASRI (Proposal No. 2016B1303). IS was supported by the DFG within the framework of the graduate school 1355 "Physics with new advanced coherent radiation sources". YF acknowledges funding by the Cluster of Excellence RESOLV (EXC 1069) funded by DFG. CS thanks the BMBF (Project 05K13PE2 within FSP-302). We thank T.~B\"uning for technical support, T.~Buslaps for support during the experiments at ESRF, M.~Brancewicz for providing the Monte Carlo code for multiple scattering calculations, and M.~Hakala for provision of the DFT data. M.~Paulus, M.~Tolan, and G.~Gr\"ubel are acknowledged for discussions and support.

\section*{Author contribution}
FL and CS designed research and wrote the beamtime proposals; FL, YF, and IS prepared the samples; FL, YF, IS, CJS, and CW performed the experiment at the ESRF; FL, YF, and ME performed the experiment at SPring-8; AP set-up and supervised the experimental procedures and ESRF; NT, MI, and YS set-up and supervised the experimental procedures at SPring-8; FL and CS analyzed and interpreted the data; FL wrote the manuscript with comments of all authors.

\providecommand*{\mcitethebibliography}{\thebibliography}
\csname @ifundefined\endcsname{endmcitethebibliography}
{\let\endmcitethebibliography\endthebibliography}{}


\begin{mcitethebibliography}{40}
	\providecommand*{\natexlab}[1]{#1}
	\providecommand*{\mciteSetBstSublistMode}[1]{}
	\providecommand*{\mciteSetBstMaxWidthForm}[2]{}
	\providecommand*{\mciteBstWouldAddEndPuncttrue}
	{\def\EndOfBibitem{\unskip.}}
	\providecommand*{\mciteBstWouldAddEndPunctfalse}
	{\let\EndOfBibitem\relax}
	\providecommand*{\mciteSetBstMidEndSepPunct}[3]{}
	\providecommand*{\mciteSetBstSublistLabelBeginEnd}[3]{}
	\providecommand*{\EndOfBibitem}{}
	\mciteSetBstSublistMode{f}
	\mciteSetBstMaxWidthForm{subitem}
	{(\emph{\alph{mcitesubitemcount}})}
	\mciteSetBstSublistLabelBeginEnd{\mcitemaxwidthsubitemform\space}
	{\relax}{\relax}
	
	\bibitem[Debenedetti(2003)]{debenedetti2003}
	P.~G. Debenedetti, \emph{J. Phys.: Condens. Matter}, 2003, \textbf{15},
	R1669\relax
	\mciteBstWouldAddEndPuncttrue
	\mciteSetBstMidEndSepPunct{\mcitedefaultmidpunct}
	{\mcitedefaultendpunct}{\mcitedefaultseppunct}\relax
	\EndOfBibitem
	\bibitem[Nilsson and Pettersson(2011)]{nilsson2011}
	A.~Nilsson and L.~Pettersson, \emph{Chem. Phys.}, 2011, \textbf{389}, 1 --
	34\relax
	\mciteBstWouldAddEndPuncttrue
	\mciteSetBstMidEndSepPunct{\mcitedefaultmidpunct}
	{\mcitedefaultendpunct}{\mcitedefaultseppunct}\relax
	\EndOfBibitem
	\bibitem[Nilsson and Pettersson(2015)]{nilsson2015}
	A.~Nilsson and L.~Pettersson, \emph{Nat. Commun.}, 2015, \textbf{6}, 8998\relax
	\mciteBstWouldAddEndPuncttrue
	\mciteSetBstMidEndSepPunct{\mcitedefaultmidpunct}
	{\mcitedefaultendpunct}{\mcitedefaultseppunct}\relax
	\EndOfBibitem
	\bibitem[Perakis \emph{et~al.}(2017)Perakis, Amann-Winkel, Lehmk{\"u}hler,
	Sprung, Mariedahl, Sellberg, Pathak, Sp{\"a}h, Cavalca,
	Schlesinger,\emph{et~al.}]{Perakis2017}
	F.~Perakis, K.~Amann-Winkel, F.~Lehmk{\"u}hler, M.~Sprung, D.~Mariedahl, J.~A.
	Sellberg, H.~Pathak, A.~Sp{\"a}h, F.~Cavalca, D.~Schlesinger \emph{et~al.},
	\emph{Proceedings of the National Academy of Sciences}, 2017,
	201705303\relax
	\mciteBstWouldAddEndPuncttrue
	\mciteSetBstMidEndSepPunct{\mcitedefaultmidpunct}
	{\mcitedefaultendpunct}{\mcitedefaultseppunct}\relax
	\EndOfBibitem
	\bibitem[Bhuyan(2002)]{Bhuyan2002}
	A.~K. Bhuyan, \emph{Biochemistry}, 2002, \textbf{41}, 13386--13394\relax
	\mciteBstWouldAddEndPuncttrue
	\mciteSetBstMidEndSepPunct{\mcitedefaultmidpunct}
	{\mcitedefaultendpunct}{\mcitedefaultseppunct}\relax
	\EndOfBibitem
	\bibitem[Ma \emph{et~al.}(2014)Ma, Pazos, and Gai]{Ma2014}
	J.~Ma, I.~M. Pazos and F.~Gai, \emph{Proceedings of the National Academy of
		Sciences}, 2014, \textbf{111}, 8476--8481\relax
	\mciteBstWouldAddEndPuncttrue
	\mciteSetBstMidEndSepPunct{\mcitedefaultmidpunct}
	{\mcitedefaultendpunct}{\mcitedefaultseppunct}\relax
	\EndOfBibitem
	\bibitem[Batchelor \emph{et~al.}(2004)Batchelor, Olteanu, Tripathy, and
	Pielak]{Batchelor2004}
	J.~D. Batchelor, A.~Olteanu, A.~Tripathy and G.~J. Pielak, \emph{J. Am. Chem.
		Soc.}, 2004, \textbf{126}, 1958--1961\relax
	\mciteBstWouldAddEndPuncttrue
	\mciteSetBstMidEndSepPunct{\mcitedefaultmidpunct}
	{\mcitedefaultendpunct}{\mcitedefaultseppunct}\relax
	\EndOfBibitem
	\bibitem[Pazos and Gai(2012)]{Pazos2012}
	I.~M. Pazos and F.~Gai, \emph{J. Phys. Chem. B}, 2012, \textbf{116},
	12473\relax
	\mciteBstWouldAddEndPuncttrue
	\mciteSetBstMidEndSepPunct{\mcitedefaultmidpunct}
	{\mcitedefaultendpunct}{\mcitedefaultseppunct}\relax
	\EndOfBibitem
	\bibitem[Knake \emph{et~al.}(2015)Knake, Schwaab, Kartaschew, and
	Havenith]{Knake2015}
	L.~Knake, G.~Schwaab, K.~Kartaschew and M.~Havenith, \emph{J. Phys. Chem. B},
	2015, \textbf{119}, 13842--13851\relax
	\mciteBstWouldAddEndPuncttrue
	\mciteSetBstMidEndSepPunct{\mcitedefaultmidpunct}
	{\mcitedefaultendpunct}{\mcitedefaultseppunct}\relax
	\EndOfBibitem
	\bibitem[Rezus and Bakker(2009)]{Rezus2009}
	Y.~Rezus and H.~Bakker, \emph{J. Phys. Chem. B}, 2009, \textbf{113},
	4038--4044\relax
	\mciteBstWouldAddEndPuncttrue
	\mciteSetBstMidEndSepPunct{\mcitedefaultmidpunct}
	{\mcitedefaultendpunct}{\mcitedefaultseppunct}\relax
	\EndOfBibitem
	\bibitem[Munroe \emph{et~al.}(2011)Munroe, Magers, and Hammer]{Munroe2011}
	K.~L. Munroe, D.~H. Magers and N.~I. Hammer, \emph{J. Phys. Chem. B}, 2011,
	\textbf{115}, 7699--7707\relax
	\mciteBstWouldAddEndPuncttrue
	\mciteSetBstMidEndSepPunct{\mcitedefaultmidpunct}
	{\mcitedefaultendpunct}{\mcitedefaultseppunct}\relax
	\EndOfBibitem
	\bibitem[Hunger \emph{et~al.}(2012)Hunger, Tielrooij, Buchner, Bonn, and
	Bakker]{Hunger2012}
	J.~Hunger, K.-J. Tielrooij, R.~Buchner, M.~Bonn and H.~J. Bakker, \emph{J.
		Phys. Chem. B}, 2012, \textbf{116}, 4783--4795\relax
	\mciteBstWouldAddEndPuncttrue
	\mciteSetBstMidEndSepPunct{\mcitedefaultmidpunct}
	{\mcitedefaultendpunct}{\mcitedefaultseppunct}\relax
	\EndOfBibitem
	\bibitem[Sahle \emph{et~al.}(2016)Sahle, Schroer, Juurinen, and
	Niskanen]{Sahle2016}
	C.~J. Sahle, M.~A. Schroer, I.~Juurinen and J.~Niskanen, \emph{Phys. Chem.
		Chem. Phys.}, 2016, \textbf{18}, 16518--16526\relax
	\mciteBstWouldAddEndPuncttrue
	\mciteSetBstMidEndSepPunct{\mcitedefaultmidpunct}
	{\mcitedefaultendpunct}{\mcitedefaultseppunct}\relax
	\EndOfBibitem
	\bibitem[Hakala \emph{et~al.}(2006)Hakala, Nyg{\aa}rd, Manninen, Huotari,
	Buslaps, Nilsson, Pettersson, and H\"{a}m\"{a}l\"{a}inen]{hakala2006}
	M.~Hakala, K.~Nyg{\aa}rd, S.~Manninen, S.~Huotari, T.~Buslaps, A.~Nilsson,
	L.~G.~M. Pettersson and K.~H\"{a}m\"{a}l\"{a}inen, \emph{J. Chem. Phys.}, 2006,
	\textbf{125}, 084504\relax
	\mciteBstWouldAddEndPuncttrue
	\mciteSetBstMidEndSepPunct{\mcitedefaultmidpunct}
	{\mcitedefaultendpunct}{\mcitedefaultseppunct}\relax
	\EndOfBibitem
	\bibitem[Rastogi \emph{et~al.}(2011)Rastogi, Ghosh, and Suresh]{Rastogi2011}
	A.~Rastogi, A.~K. Ghosh and S.~Suresh, \emph{Thermodynamics-Physical Chemistry
		of Aqueous Systems}, InTech, 2011\relax
	\mciteBstWouldAddEndPuncttrue
	\mciteSetBstMidEndSepPunct{\mcitedefaultmidpunct}
	{\mcitedefaultendpunct}{\mcitedefaultseppunct}\relax
	\EndOfBibitem
	\bibitem[Sahle \emph{et~al.}(2013)Sahle, Sternemann, Schmidt, Lehtola, Jahn,
	Simonelli, Huotari, Hakala, Pylkk\"{a}nen, Nyrow, Mende, Tolan, H\"{a}m\"{a}l\"{a}inen, and
	Wilke]{sahle2013}
	C.~J. Sahle, C.~Sternemann, C.~Schmidt, S.~Lehtola, S.~Jahn, L.~Simonelli,
	S.~Huotari, M.~Hakala, T.~Pylkk\"{a}nen, A.~Nyrow, K.~Mende, M.~Tolan,
	K.~H\"{a}m\"{a}l\"{a}inen and M.~Wilke, \emph{PNAS}, 2013, \textbf{110}, 6301--6306\relax
	\mciteBstWouldAddEndPuncttrue
	\mciteSetBstMidEndSepPunct{\mcitedefaultmidpunct}
	{\mcitedefaultendpunct}{\mcitedefaultseppunct}\relax
	\EndOfBibitem
	\bibitem[Sit \emph{et~al.}(2007)Sit, Bellin, Barbiellini, Testemale, Hazemann,
	Buslaps, Marzari, and Shukla]{sit2007}
	P.~H.-L. Sit, C.~Bellin, B.~Barbiellini, D.~Testemale, J.-L. Hazemann,
	T.~Buslaps, N.~Marzari and A.~Shukla, \emph{Phys. Rev. B}, 2007, \textbf{76},
	245413\relax
	\mciteBstWouldAddEndPuncttrue
	\mciteSetBstMidEndSepPunct{\mcitedefaultmidpunct}
	{\mcitedefaultendpunct}{\mcitedefaultseppunct}\relax
	\EndOfBibitem
	\bibitem[Lehmk\"uhler \emph{et~al.}(2011)Lehmk\"uhler, Sakko, Steinke,
	Sternemann, Hakala, Sahle, Buslaps, Simonelli, Galambosi, Paulus,
	Pylkk\"anen, Tolan, and H\"am\"al\"ainen]{lehmkuehler2011}
	F.~Lehmk\"uhler, A.~Sakko, I.~Steinke, C.~Sternemann, M.~Hakala, C.~J. Sahle,
	T.~Buslaps, L.~Simonelli, S.~Galambosi, M.~Paulus, T.~Pylkk\"anen, M.~Tolan
	and K.~H\"am\"al\"ainen, \emph{J. Phys. Chem. C}, 2011, \textbf{115},
	21009--21015\relax
	\mciteBstWouldAddEndPuncttrue
	\mciteSetBstMidEndSepPunct{\mcitedefaultmidpunct}
	{\mcitedefaultendpunct}{\mcitedefaultseppunct}\relax
	\EndOfBibitem
	\bibitem[Isaacs \emph{et~al.}(1999)Isaacs, Shukla, Platzman, Hamann,
	Barbiellini, and Tulk]{isaacs1999}
	E.~D. Isaacs, A.~Shukla, P.~M. Platzman, D.~R. Hamann, B.~Barbiellini and C.~A.
	Tulk, \emph{Phys. Rev. Lett.}, 1999, \textbf{82}, 600--603\relax
	\mciteBstWouldAddEndPuncttrue
	\mciteSetBstMidEndSepPunct{\mcitedefaultmidpunct}
	{\mcitedefaultendpunct}{\mcitedefaultseppunct}\relax
	\EndOfBibitem
	\bibitem[Bellin \emph{et~al.}(2011)Bellin, Barbiellini, Klotz, Buslaps, Rousse,
	Str\"assle, and Shukla]{bellin2011}
	C.~Bellin, B.~Barbiellini, S.~Klotz, T.~Buslaps, G.~Rousse, T.~Str\"assle and
	A.~Shukla, \emph{Phys. Rev. B}, 2011, \textbf{83}, 094117\relax
	\mciteBstWouldAddEndPuncttrue
	\mciteSetBstMidEndSepPunct{\mcitedefaultmidpunct}
	{\mcitedefaultendpunct}{\mcitedefaultseppunct}\relax
	\EndOfBibitem
	\bibitem[Juurinen \emph{et~al.}(2011)Juurinen, Nakahara, Ando, Nishiumi, Seta,
	Yoshida, Morinaga, Itou, Ninomiya, Sakurai, Salonen, Nordlund,
	H\"am\"al\"ainen, and Hakala]{juurinen2011}
	I.~Juurinen, K.~Nakahara, N.~Ando, T.~Nishiumi, H.~Seta, N.~Yoshida,
	T.~Morinaga, M.~Itou, T.~Ninomiya, Y.~Sakurai, E.~Salonen, K.~Nordlund,
	K.~H\"am\"al\"ainen and M.~Hakala, \emph{Phys. Rev. Lett.}, 2011,
	\textbf{107}, 197401\relax
	\mciteBstWouldAddEndPuncttrue
	\mciteSetBstMidEndSepPunct{\mcitedefaultmidpunct}
	{\mcitedefaultendpunct}{\mcitedefaultseppunct}\relax
	\EndOfBibitem
	\bibitem[Juurinen \emph{et~al.}(2013)Juurinen, Pylkk\"anen, Ruotsalainen,
	Sahle, Monaco, H\"{a}m\"{a}l\"{a}inen, Huotari, and Hakala]{juurinen2013}
	I.~Juurinen, T.~Pylkk\"anen, K.~O. Ruotsalainen, C.~J. Sahle, G.~Monaco,
	K.~H\"{a}m\"{a}l\"{a}inen, S.~Huotari and M.~Hakala, \emph{J. Phys. Chem. B}, 2013,
	\textbf{117}, 16506--16511\relax
	\mciteBstWouldAddEndPuncttrue
	\mciteSetBstMidEndSepPunct{\mcitedefaultmidpunct}
	{\mcitedefaultendpunct}{\mcitedefaultseppunct}\relax
	\EndOfBibitem
	\bibitem[Juurinen \emph{et~al.}(2014)Juurinen, Pylkk\"anen, Sahle, Simonelli,
	H\"{a}m\"{a}l\"{a}inen, Huotari, and Hakala]{juurinen2014b}
	I.~Juurinen, T.~Pylkk\"anen, C.~J. Sahle, L.~Simonelli, K.~H\"{a}m\"{a}l\"{a}inen,
	S.~Huotari and M.~Hakala, \emph{J. Phys. Chem. B}, 2014, \textbf{118},
	8750--8755\relax
	\mciteBstWouldAddEndPuncttrue
	\mciteSetBstMidEndSepPunct{\mcitedefaultmidpunct}
	{\mcitedefaultendpunct}{\mcitedefaultseppunct}\relax
	\EndOfBibitem
	\bibitem[Lehmk{\"{u}}hler \emph{et~al.}(2016)Lehmk{\"{u}}hler, Forov,
	B{\"{u}}ning, Sahle, Steinke, Julius, Buslaps, Tolan, Hakala, and
	Sternemann]{lehmkuehler2016}
	F.~Lehmk{\"{u}}hler, Y.~Forov, T.~B{\"{u}}ning, C.~J. Sahle, I.~Steinke,
	K.~Julius, T.~Buslaps, M.~Tolan, M.~Hakala and C.~Sternemann, \emph{Phys.
		Chem. Chem. Phys.}, 2016, \textbf{18}, 6925--6930\relax
	\mciteBstWouldAddEndPuncttrue
	\mciteSetBstMidEndSepPunct{\mcitedefaultmidpunct}
	{\mcitedefaultendpunct}{\mcitedefaultseppunct}\relax
	\EndOfBibitem
	\bibitem[Cooper(2004)]{Cooper2004}
	M.~Cooper, \emph{X-ray Compton scattering}, Oxford University Press on Demand,
	2004\relax
	\mciteBstWouldAddEndPuncttrue
	\mciteSetBstMidEndSepPunct{\mcitedefaultmidpunct}
	{\mcitedefaultendpunct}{\mcitedefaultseppunct}\relax
	\EndOfBibitem
	\bibitem[Sch\"{u}lke(2007)]{schuelke}
	W.~Sch\"{u}lke, \emph{Electron Dynamics by Inelastic X-ray Scattering}, Oxford
	University Press, Oxford, 2007\relax
	\mciteBstWouldAddEndPuncttrue
	\mciteSetBstMidEndSepPunct{\mcitedefaultmidpunct}
	{\mcitedefaultendpunct}{\mcitedefaultseppunct}\relax
	\EndOfBibitem
	\bibitem[Koskelo \emph{et~al.}(2014)Koskelo, Juurinen, Ruotsalainen, McGrath,
	Kuo, Lehtola, Galambosi, H\"{a}m\"{a}l\"{a}inen, Huotari, and Hakala]{koskelo}
	J.~Koskelo, I.~Juurinen, K.~O. Ruotsalainen, M.~J. McGrath, I.-F. Kuo,
	S.~Lehtola, S.~Galambosi, K.~H\"{a}m\"{a}l\"{a}inen, S.~Huotari and M.~Hakala, \emph{J.
		Chem. Phys.}, 2014, \textbf{141}, 244505\relax
	\mciteBstWouldAddEndPuncttrue
	\mciteSetBstMidEndSepPunct{\mcitedefaultmidpunct}
	{\mcitedefaultendpunct}{\mcitedefaultseppunct}\relax
	\EndOfBibitem
	\bibitem[Juurinen \emph{et~al.}(2014)Juurinen, Galambosi, Anghelescu-Hakala,
	Koskelo, Honkim\"{a}ki, H\"{a}m\"{a}l\"{a}inen, Huotari, and Hakala]{juurinen2014}
	I.~Juurinen, S.~Galambosi, A.~G. Anghelescu-Hakala, J.~Koskelo, V.~Honkim\"{a}ki,
	K.~H\"{a}m\"{a}l\"{a}inen, S.~Huotari and M.~Hakala, \emph{J. Phys. Chem. B}, 2014,
	\textbf{118}, 5518--5523\relax
	\mciteBstWouldAddEndPuncttrue
	\mciteSetBstMidEndSepPunct{\mcitedefaultmidpunct}
	{\mcitedefaultendpunct}{\mcitedefaultseppunct}\relax
	\EndOfBibitem
	\bibitem[Hakala \emph{et~al.}(2006)Hakala, Nyg\aa{}rd, Manninen, Pettersson,
	and H\"am\"al\"ainen]{hakala2006b}
	M.~Hakala, K.~Nyg\aa{}rd, S.~Manninen, L.~G.~M. Pettersson and
	K.~H\"am\"al\"ainen, \emph{Phys. Rev. B}, 2006, \textbf{73}, 035432\relax
	\mciteBstWouldAddEndPuncttrue
	\mciteSetBstMidEndSepPunct{\mcitedefaultmidpunct}
	{\mcitedefaultendpunct}{\mcitedefaultseppunct}\relax
	\EndOfBibitem
	\bibitem[Nyg{\aa}rd \emph{et~al.}(2007)Nyg{\aa}rd, Hakala, Pylkk\"{a}nen, Manninen,
	Buslaps, Itou, Andrejczuk, Sakurai, Odelius, and H\"{a}m\"{a}l\"{a}inen]{nygard2007}
	K.~Nyg{\aa}rd, M.~Hakala, T.~Pylkk\"{a}nen, S.~Manninen, T.~Buslaps, M.~Itou,
	A.~Andrejczuk, Y.~Sakurai, M.~Odelius and K.~H\"{a}m\"{a}l\"{a}inen, \emph{J. Chem.
		Phys.}, 2007, \textbf{126}, 154508\relax
	\mciteBstWouldAddEndPuncttrue
	\mciteSetBstMidEndSepPunct{\mcitedefaultmidpunct}
	{\mcitedefaultendpunct}{\mcitedefaultseppunct}\relax
	\EndOfBibitem
	\bibitem[Reiter \emph{et~al.}(2013)Reiter, Deb, Sakurai, Itou, Krishnan, and
	Paddison]{reiter2013}
	G.~F. Reiter, A.~Deb, Y.~Sakurai, M.~Itou, V.~G. Krishnan and S.~J. Paddison,
	\emph{Phys. Rev. Lett.}, 2013, \textbf{111}, 036803\relax
	\mciteBstWouldAddEndPuncttrue
	\mciteSetBstMidEndSepPunct{\mcitedefaultmidpunct}
	{\mcitedefaultendpunct}{\mcitedefaultseppunct}\relax
	\EndOfBibitem
	\bibitem[Reiter \emph{et~al.}(2016)Reiter, Deb, Sakurai, Itou, and
	Kolesnikov]{reiter2016}
	G.~F. Reiter, A.~Deb, Y.~Sakurai, M.~Itou and A.~I. Kolesnikov, \emph{J. Phys.
		Chem. Lett.}, 2016, \textbf{7}, 4433--4437\relax
	\mciteBstWouldAddEndPuncttrue
	\mciteSetBstMidEndSepPunct{\mcitedefaultmidpunct}
	{\mcitedefaultendpunct}{\mcitedefaultseppunct}\relax
	\EndOfBibitem
	\bibitem[Ragot \emph{et~al.}(2002)Ragot, Gillet, and Becker]{ragot2002}
	S.~Ragot, J.-M. Gillet and P.~J. Becker, \emph{Phys. Rev. B}, 2002,
	\textbf{65}, 235115\relax
	\mciteBstWouldAddEndPuncttrue
	\mciteSetBstMidEndSepPunct{\mcitedefaultmidpunct}
	{\mcitedefaultendpunct}{\mcitedefaultseppunct}\relax
	\EndOfBibitem
	\bibitem[Nyg\aa{}rd \emph{et~al.}(2007)Nyg\aa{}rd, Hakala, Manninen, Itou,
	Sakurai, and H\"am\"al\"ainen]{nygardprl}
	K.~Nyg\aa{}rd, M.~Hakala, S.~Manninen, M.~Itou, Y.~Sakurai and
	K.~H\"am\"al\"ainen, \emph{Phys. Rev. Lett.}, 2007, \textbf{99}, 197401\relax
	\mciteBstWouldAddEndPuncttrue
	\mciteSetBstMidEndSepPunct{\mcitedefaultmidpunct}
	{\mcitedefaultendpunct}{\mcitedefaultseppunct}\relax
	\EndOfBibitem
	\bibitem[Sternemann \emph{et~al.}(2006)Sternemann, Huotari, Hakala, Paulus,
	Volmer, Gutt, Buslaps, Hiraoka, Klug, H\"am\"al\"ainen, Tolan, and
	Tse]{sternemann2006}
	C.~Sternemann, S.~Huotari, M.~Hakala, M.~Paulus, M.~Volmer, C.~Gutt,
	T.~Buslaps, N.~Hiraoka, D.~D. Klug, K.~H\"am\"al\"ainen, M.~Tolan and J.~S.
	Tse, \emph{Phys. Rev. B}, 2006, \textbf{73}, 195104\relax
	\mciteBstWouldAddEndPuncttrue
	\mciteSetBstMidEndSepPunct{\mcitedefaultmidpunct}
	{\mcitedefaultendpunct}{\mcitedefaultseppunct}\relax
	\EndOfBibitem
	\bibitem[Hakala \emph{et~al.}(2009)Hakala, Nyg\aa{}rd, Vaara, Itou, Sakurai,
	and H\"{a}m\"{a}l\"{a}inen]{hakala2009}
	M.~Hakala, K.~Nyg\aa{}rd, J.~Vaara, M.~Itou, Y.~Sakurai and K.~H\"{a}m\"{a}l\"{a}inen,
	\emph{J. Chem. Phys.}, 2009, \textbf{130}, 034506\relax
	\mciteBstWouldAddEndPuncttrue
	\mciteSetBstMidEndSepPunct{\mcitedefaultmidpunct}
	{\mcitedefaultendpunct}{\mcitedefaultseppunct}\relax
	\EndOfBibitem
	\bibitem[Lehmk\"uhler \emph{et~al.}(2010)Lehmk\"uhler, Sakko, Sternemann,
	Hakala, Nyg\aa{}rd, Sahle, Galambosi, Steinke, Tiemeyer, Nyrow, Buslaps,
	Pontoni, Tolan, and H\"am\"al\"ainen]{lehmkuehler2010}
	F.~Lehmk\"uhler, A.~Sakko, C.~Sternemann, M.~Hakala, K.~Nyg\aa{}rd, C.~J.
	Sahle, S.~Galambosi, I.~Steinke, S.~Tiemeyer, A.~Nyrow, T.~Buslaps,
	D.~Pontoni, M.~Tolan and K.~H\"am\"al\"ainen, \emph{J. Phys. Chem. Lett.},
	2010, \textbf{1}, 2832--2836\relax
	\mciteBstWouldAddEndPuncttrue
	\mciteSetBstMidEndSepPunct{\mcitedefaultmidpunct}
	{\mcitedefaultendpunct}{\mcitedefaultseppunct}\relax
	\EndOfBibitem
	\bibitem[Hiraoka \emph{et~al.}(2005)Hiraoka, Buslaps, Honkim{\"{a}}ki, and
	Suortti]{Hiraoka2005}
	N.~Hiraoka, T.~Buslaps, V.~Honkim{\"{a}}ki and P.~Suortti, \emph{J. Synchrotron
		Rad.}, 2005, \textbf{12}, 670--674\relax
	\mciteBstWouldAddEndPuncttrue
	\mciteSetBstMidEndSepPunct{\mcitedefaultmidpunct}
	{\mcitedefaultendpunct}{\mcitedefaultseppunct}\relax
	\EndOfBibitem
	\bibitem[Hiraoka \emph{et~al.}(2001)Hiraoka, Itou, Ohata, Mizumaki, Sakurai,
	and Sakai]{Hiraoka2001}
	N.~Hiraoka, M.~Itou, T.~Ohata, M.~Mizumaki, Y.~Sakurai and N.~Sakai, \emph{J.
		Synchrotron Rad.}, 2001, \textbf{8}, 26--32\relax
	\mciteBstWouldAddEndPuncttrue
	\mciteSetBstMidEndSepPunct{\mcitedefaultmidpunct}
	{\mcitedefaultendpunct}{\mcitedefaultseppunct}\relax
	\EndOfBibitem
	\bibitem[Brancewicz \emph{et~al.}(2016)Brancewicz, Itou, and
	Sakurai]{montecarlo}
	M.~Brancewicz, M.~Itou and Y.~Sakurai, \emph{J. Synchrotron Radiat.}, 2016,
	\textbf{23}, 244--252\relax
	\mciteBstWouldAddEndPuncttrue
	\mciteSetBstMidEndSepPunct{\mcitedefaultmidpunct}
	{\mcitedefaultendpunct}{\mcitedefaultseppunct}\relax
	\EndOfBibitem
\end{mcitethebibliography}
\end{document}